\documentstyle[preprint,aps,epsfig]{revtex}

\tighten
\draft

\preprint{\vbox{\hbox{PNU-NTG-04/99} \hbox{RUB-TPII-15/99}}}
\begin{document}
\title{Hyperon semileptonic decays and quark spin content of the proton}
\author{Hyun-Chul Kim$^a$,
Micha{\l} Prasza{\l}owicz$^b$, and Klaus Goeke$^c$}
\address{~\\
$^a$ Department of Physics, Pusan National University,\\
Pusan 609-735, Republic of Korea \\
$^b$ Institute of Physics, Jagellonian University, \\
ul. Reymonta 4, 30-059 Krak{\'o}w, Poland \\
$^c$ Institute for Theoretical Physics II, Ruhr-University Bochum, \\
D-44780 Bochum, Germany}
\date{Januray, 2000}
\maketitle

\begin{abstract}
We investigate the hyperon semileptonic decays and the quark spin
content of the proton $\Delta \Sigma$ taking into account flavor SU(3)
symmetry breaking. Symmetry breaking is implemented with the help of
the chiral quark-soliton model in an approach, in which the dynamical
parameters are fixed by the experimental data for six hyperon
semileptonic decay constants. As a result  we predict the unmeasured
decay constants, particularly for $\Xi^0 \rightarrow \Sigma^+$, which
will be soon measured and examine the effect of the SU(3) symmetry
breaking on the spin content $\Delta \Sigma $ of the proton.
Unfortunately large experimental errors of $\Xi^-$ decays propagate in
our analysis making $\Delta \Sigma$ and $\Delta s$ practically
undetermined.  We conclude that statements concerning the values of
these two quantities, which are based on the exact SU(3) symmetry, are
premature. We stress that the meaningful results can be obtained only
if the experimental errors for the $\Xi$ decays are reduced.
\end{abstract}


\section{Introduction}

Since the European Muon Collaboration (EMC) measured the first moment
$I_p^{{\rm EMC}}=0.112$ (at $Q^2$=3~(GeV/c)$^2$) of the proton spin
structure function $g^p_{1}$~\cite{EMC}, there has been a great deal of
discussion about the spin content of the proton. An  immediate and
unexpected consequence of the EMC measurement was that the quark
contribution to the spin of the proton was very small ($\Delta \Sigma
\approx 0$).  A series of following experiments~\cite{SMC,E143,SMC_d}
confirmed the EMC measurement, giving, however a somewhat larger, but
still small value for $\Delta \Sigma$.

This result is in contradiction with expectations based on the naive,
nonrelativistic quark model, supplemented by the assumption that the
contribution of strange quarks to $I_p$ was zero ($\Delta s=0$)~\cite
{EllisJaffe}. The EMC measurements required $\Delta s \ne 0$ and
relatively large. These two results: $\Delta \Sigma \approx 0$ and
$\Delta s \ne 0$ are often referred to as {\em spin crisis}. Let us
shortly summarize how the crisis arises.

Theoretical analysis of recent measurements~\cite{EllKar} indicates
that the $I_p$ is equal to:
\begin{equation}
I_p(Q^2=3~({\rm GeV}/c)^2)=0.124\pm 0.011.  \label{Ip}
\end{equation}
On the other hand the $I_p$ is related to the integrated
polarized quark densities:
\begin{equation}
I_p=\frac 1{18}(4\Delta u+\Delta d+\Delta s)
\left( 1-\frac{\alpha _{{\rm s}}} \pi +\ldots \right) . \label{Ip2}
\end{equation}
Here for simplicity we neglect higher orders and higher twist
contributions.  Comparing Eq.(\ref{Ip}) with Eq.(\ref{Ip2}) and
assuming $\alpha _{{\rm s} }(Q^2=3~({\rm
GeV}/c)^2)=0.4$~\cite{KarLip}, we get immediately:
\begin{equation}
\Gamma _p\equiv 4\Delta u+\Delta d+\Delta s=2.56\pm 0.23\,.
\label{Gampval}
\end{equation}
Let us quote here for completeness the experimental value for the
neutron~\cite{EllKar}:
\begin{equation}
\Gamma _n\equiv 4\Delta d+\Delta u+\Delta s=-0.928\pm 0.186\,.
\label{Gamnval}
\end{equation}
With this definition of $\Gamma _n$ the Bjorken sum rule is automatically
satisfied.

Integrated polarized quark densities $\Delta q$ can be in principle
extracted from the hyperon semileptonic decays. It is customary to
assume SU(3) symmetry to analyze these decays. Then all decay
amplitudes are given in terms of two reduced matrix elements $F$ and
$D$. For example:
\[
A_1({\rm n}    \rightarrow {\rm p})=F+D\,,~~~~
A_4(\Sigma ^{-}\rightarrow {\rm n})=F-D\,.
\]
Here by $A_i$ we denote the ratios of axial-vector to vector coupling
constants $g_1/f_1$ for semileptonic decays as displayed in Table I.
Taking for these decays experimental values (see Table I) one gets
$F=0.46$ and $D=0.80$. The matrix elements of diagonal operators $%
\lambda _3$ and $\lambda _8$ (called $g_{{\rm A}}^{(3)}$ and $g_{{\rm A}%
}^{(8)}$ respectively), which define integrated quark densities
$\Delta q$, can be also
expressed in terms of $F$ and $D$:
\begin{eqnarray}
g_{{\rm A}}^{(3)}\equiv \Delta u-\Delta d &=&F+D\,,  \nonumber \\
g_{{\rm A}}^{(8)}\equiv \frac 1{\sqrt{3}}\left( \Delta u+\Delta d-2\Delta
s\right) &=&\frac 1{\sqrt{3}}\left( 3F-D\right) \,.
\end{eqnarray}
Using the values of $F$ and $D$ obtained from the neutron and $\Sigma ^{-}$
decays together with Eq.(\ref{Gampval}) we get: $\Delta u=0.79$, $\Delta
d=-0.47$, and $\Delta s=-0.13$. Defining the quark content of the proton's
spin:
\begin{equation}
\Delta \Sigma =\Delta u+\Delta d+\Delta s  \label{Sigma}
\end{equation}
we obtain $\Delta \Sigma =0.19$. Had we used for $I_p$ the result of the
first EMC measurement $I_p^{{\rm EMC}}=0.112$, we would get even smaller
value: $\Delta \Sigma =0.07$.

Although quite often used, the above derivation of $\Delta \Sigma$ has,
however, one serious flaw.
Namely we could equally well use some other decays to extract $F$ and $D$.
For example using:
\begin{eqnarray}
A_4(\Sigma^- \rightarrow {\rm n}) = F-D \, , &~~~~ & A_5(\Xi^- \rightarrow
\Lambda) = F - \frac{D}{3} \, ,  \label{ch2}
\end{eqnarray}
together with the experimental data for these decays (see Table I) and
experimental value for $\Gamma_p$, Eq.(\ref{Gampval}),
we would get $F=0.55$ and $D=0.89$,
yielding $\Delta \Sigma=0.02$ -- almost ten times less than our previous
value. It is the breaking of the SU(3) symmetry, which is responsible for
this discrepancy. Although the symmetry breaking in hyperon decays
themselves is not that large, {\em i.e.} it amounts to no more than
10 \%, the effect of the symmetry breaking on $\Delta \Sigma$, or
integrated quark density $\Delta s$, is much stronger.

There are 6 measured semileptonic hyperon decays, so that the number of
combinations which one can form to extract $F$ and $D$ is 14 (actually 15,
but two conditions are linearly dependent). Taking these 14 combinations
into account and Eq.(\ref{Gampval}) we get the following values for $\Delta
u=0.75\rightarrow 0.85$, $\Delta d=-0.39\rightarrow -0.58$ and
$\Delta s=-0.05\rightarrow -0.25 $,
which in turn give $\Delta \Sigma =0.02 \rightarrow 0.30$. These are the
uncertainties of the {\em central values} due to the theoretical error
caused by using SU(3) symmetry to describe the hyperon decays. They are
further increased by the experimental errors of all individual decays and
the one of $\Gamma _p$.

The authors of Ref.\cite{EhrSch} made similar
observation trying to fit the variation of $F$ and $D$ for various
decays with one parameter related to $m_{\rm s}$. Assuming further
$\Delta s=0$ they were able to fit experimental data for
$I_{p,n,deuter}$ with satisfactory accuracy.

Similarly in Refs.\cite{Lipkin,LichtLip} a simple quark model has been
proposed to describe the symmetry breaking in the hyperon decays.
It has been observed that with the increase of the symmetry breaking
parameter the value of $\Delta s$ increased, while $\Delta \Sigma$
stayed almost unchanged.

Semileptonic decays and $\Delta \Sigma$ have been also investigated
within the SU(3) Skyrme Model
\cite{BroEllKar}\nocite{PSW1}-\cite{PSW2}, where $\Delta \Sigma =0$
irrespectively of the symmetry breaking. Symmetry breaking influences
only $\Delta s$ \cite{PSW1,PSW2}. In this respect our analysis gives a
similar result: although $\Delta \Sigma \ne 0$ it depends very weakly
on $m_{\rm s}$.

It is virtually impossible to analyze the symmetry breaking in weak
decays without resorting to some specific model \cite{KarLip}. In this
paper we will implement the symmetry breaking for the hyperon decays
using the Chiral Quark-Soliton Model ($\chi $QSM, see Ref.\cite{review}
for review).  This model has proven to give satisfactory description of
the axial-vector properties of hyperons
\cite{BloPraGo}\nocite{BPG,Wakaspin}--\cite{KimPoPraGo}.
It describes  the baryons as solitons rotating adiabatically in flavor
space.  Thus it provides a link between the matrix elements of the
octet of the axial-vector currents, responsible for hyperon decays, and
the matrix elements of the singlet axial-vector current, in our
normalization equal to $\Delta \Sigma$.  In the present work we will
study the relation between the semileptonic decays and integrated
polarized quark distributions, with the help of the $\chi$QSM.  However
we will use only the collective Hamiltonian of the flavor rotational
degrees of freedom including the corrections linear in  the strange
quark mass $m_{\rm s}$.  The dynamical quantities in this Hamiltonian,
certain moments of inertia calculable within the model \cite{BloPraGo},
are not calculated but treated as free parameters.  By adjusting them
to the experimentally known semileptonic decays we allow for maximal
phenomenological input and minimal model dependence.  In
Ref.\cite{KimPraGo,strange} we have already studied the magnetic
moments of the octet and decuplet in this way.

Such an approach -- introduced to our knowledge for the first time by
Adkins and Nappi \cite{AdNap} in the context of the Skyrme model -- can
be viewed from two perspectives. Firstly, it can be considered as a QCD
motivated tool to analyze and clasify (in terms of powers of $m_{\rm
s}$  and $1/N_{\rm c}$) the symmetry breaking terms for a given
observable. For nontrivial operators such as magnetic moments or axial
form factors a general analysis, without refering to some specific
model, is often virtually impossible.  Secondly, it also provides
information for the model builders. It tells us what are the best
predictions the model can ever produce. Indeed, model calculations are
not as unique as one might think: they depend on adopted
regularization, cutoff parameters, constituent quark mass.  Moreover in
the SU(3) version of the $\chi$QSM the quantization ambiguity appears
\cite{paradox}. So if the ``model independent'' analysis would have
failed to describie the data, that would mean that the model did not
corretly include all necessary physics relevant for a given observable.
On the other hand the success of such an analysis gives a strong hint
for the model builders that the model is correct and worth exploring.
In fact this concerns all the hedgehog models which would give the
collective structure identical to the one of the $\chi$QSM.

As far as the symmetry breaking is concerned, our results are identical
to the ones obtained in Refs.\cite{Man} within large $N_{{\rm c}}$ QCD.
Indeed, the $\chi $QSM is a specific realization of the large $N_{{\rm
c}}$ limit. The new ingredient of our analysis is the model formula for
the singlet axial-vector constant $g_{{\rm A}}^{(0)}$, which we use to
calculate quantities relevant for the polarized high energy
experiments. In the $\chi$QSM one can define two interesting limits
\cite{DiaMog,limit,charge} in which
the soliton size is artificially changed either to zero (so called
quark-model limit), or to $%
\infty $ (Skyrme limit). In these two limiting cases one recovers the
well known results: 1) $g_{{\rm A}}^{(0)}=1$ in the quark model limit
and 2) $g_{{\rm A}}^{(0)}/g_{{\rm A}}^{(3)}\rightarrow 0$ in the Skyrme
limit. This concerns not only the axial couplings; it is often said
that the $\chi$QSM {\em interpolates} between the quark model and the
Skyrme model.  Also these simple qualitative features make us believe
that the model correctly describes physics essential for the
axial-vector properties of the nucleon.

The Skyrme limit of the $\chi$QSM can be also defined as the limit in
which the constituent quark mass $M \rightarrow \infty$. The explicit
{\em interpolating} features of the SU(2) version of the model in this
limit have been discussed numerically in Ref.\cite{WaYo}.

As we will see, in the $\chi $QSM in the chiral limit we can express
the singlet axial-vector coupling through $F$ and $D$: $g_{{\rm
A}}^{(0)}=9F-5D$.  We see that the value of $g_{{\rm A}}^{(0)}$ is very
sensitive to small variations of $F$ and $D$, since it is a difference
of the two, with relatively large multiplicators.  Indeed, for the 14
fits mentioned above (where as the input we use {\em only} semileptonic
decays plus model formula for $g_{{\rm A}}^{(0)}$) the central value
for $g_{{\rm A}}^{(0)}$ varies between $-0.25$ to approximately 1.  So
despite the fact that semileptonic decays are relatively well described
by the model in the chiral limit, the singlet axial-vector coupling is
basically undetermined. This is a clear signal of the importance of the
symmetry breaking for this quantity.

One could argue that this kind of behavior is just an artifact of the
$\chi$QSM.  However, the scenario of a rotating soliton (which is by
the way used also in the Skyrme-type models) is very plausible and
cannot be {\em a priori} discarded on the basis of first principles.
The $\chi$QSM is a particular realization of this scenario and we use
it as a tool to investigate the sensitivity of the singlet axial
current to the symmetry breaking effects in hyperon decays.  In fact
conclusions similar to ours have been obtained in chiral perturbation
theory in Ref.\cite{SavWal}.

As a result of our present analysis we will give predictions for the
semileptonic decays not yet measured. More importantly, we will show
that the symmetry breaking effects cannot be neglected in the analysis
of the quark contribution to the spin of the proton.  In other words,
linking low energy data with high energy polarized experiments is
meaningful only if the SU(3) breaking is taken into account.  We will
furthermore show which semileptonic decays should be measured more
accurately in order to reduce  the experimental errors for $\Delta
\Sigma$ and $\Delta s$.

The paper is organized as follows: in the next Section we will shortly
recapitulate the formalism of the $\chi$QSM needed for the calculation
of semileptonic hyperon decays. In Section III we will discuss
quantities relevant for the polarized parton distribution. Finally in
Section IV we will draw conclusions. Formulae used to calculate hyperon
decays and axial-vector constants are collected in the Appendix.

\section{Hyperon decays in the Chrial Quark Soliton Model}

The transition matrix elements of the hadronic axial-vector current
$\langle B_2 | A_\mu^X | B_1\rangle$ can be expressed in terms of three
independent form factors:
\begin{equation}
\langle B_2| A_\mu^X |B_1\rangle\;=\;\bar{u}_{B_2} (p_2)
\left[ \left\{g_1^{B_1\rightarrow B_2} (q^2) \gamma_\mu -
\frac{i g_2^{B_1\rightarrow B_2} (q^2)}{M_1} \sigma_{\mu\nu} q^\nu
+ \frac{g_3^{B_1\rightarrow B_2}
(q^2)}{M_1} q_\mu\right\}\gamma_5 \right] u_{B_1} (p_1),
\end{equation}
where the axial-vector current is defined as
\begin{equation}
A_{\mu}^X\;=\; \bar{\psi}(x) \gamma_\mu \gamma_5 \lambda_X \psi (x)
\label{Eq:current}
\end{equation}
with $X= \frac12 (1 \pm i 2)$ for strangeness conserving $\Delta S = 0$
currents and $X=\frac12 (4 \pm i 5) $ for $|\Delta S| = 1 $. Similar
expressions hold for the hadronic vector current, where the $g_i$ are
replaced by $f_i$ ($i=1,2,3$) and $\gamma_5$ by ${\bf 1}$.

The $q^2=-Q^2$ stands for the square of the momentum transfer
$q=p_2-p_1$.  The form factors $g_i$ are real quantities depending only
on the square of the momentum transfer in the case of $CP$-invariant
processes. We can safely neglect $g_3$ for the reason that on account
of $q_\mu $ its contribution to the decay rate is proportional to the
ratio $\frac{m_l^2}{M_1^2}\ll 1$, where $m_l$ represents the mass of
the lepton ($e$ or $\mu $) in the final state and $M_1$ that of the
baryon in the initial state .

The form factor $g_2$ is equal to 0 in the chiral limit. It gets the
first nonvanishing contribution in the linear order in $m_{\text{s}}$.
The inclusion of this effect in the discussion of the hyperon decays
would require reanalyzing the experimental data, which is beyond the
scope of this paper. However, the model calculations show that the
$m_{\text{s}}$ contribution to $g_2$ enters with relatively small
numerical coefficient, which means that the numerical error due to the
neglect of $g_2$ in the full fledged analysis of the hyperon decays is
small.

It is already well known how to treat hadronic matrix elements such as
$\langle B_2|A_\mu ^X|B_1\rangle $ within the $\chi $QSM (see for
example \cite{review} and references therein.). Taking into account the
$1/N_c$ rotational and $m_{{\rm s}}$ corrections, we can write the
resulting axial-vector constants $g_1^{B_1\rightarrow B_2}(0)$ in the
following form%
\footnote{%
In the following we will assume that the baryons involved have $S_3 =
\frac12 $.}:
\begin{eqnarray}
g_1^{(B_1\rightarrow B_2)} &=&a_1\langle B_2|D_{X3}^{(8)}|B_1\rangle
\;+\;a_2d_{pq3}\langle B_2|D_{Xp}^{(8)}\,\hat{S}_q|B_1\rangle \;
+\;\frac{a_3%
}{\sqrt{3}}\langle B_2|D_{X8}^{(8)}\,\hat{S}_3|B_1\rangle  \nonumber \\
&+&m_s\left[ \frac{a_4}{\sqrt{3}}d_{pq3}\langle
B_2|D_{Xp}^{(8)}\,D_{8q}^{(8)}|B_1\rangle +a_5\langle B_2|\left(
D_{X3}^{(8)}\,D_{88}^{(8)}+D_{X8}^{(8)}\,D_{83}^{(8)}\right) |B_1\rangle
\right.  \nonumber \\
&+&\left. a_6\langle B_2|\left(
D_{X3}^{(8)}\,D_{88}^{(8)}-D_{X8}^{(8)}\,D_{83}^{(8)}\right) |B_1\rangle
\right] .  \label{Eq:g1}
\end{eqnarray}
$\hat{S}_q$ ($\hat{S}_3$) stand for the $q$-th (third)
component of the spin operator of the baryons. The $D_{ab}^{({\cal
R})}$ denote the SU(3) Wigner matrices in representation ${\cal R}$.
The $a_i$ denote parameters depending on the specific dynamics of the
chiral soliton model. Their explicit form in terms of a Goldstone mean
field can be found in Ref.~\cite{BloPraGo}.  As mentioned already, in
the present approach we will not calculate this mean field but treat
$a_i$ as free parameters to be adjusted to experimentally known
semileptonic hyperon decays.

Because of the SU(3) symmetry breaking due to the strange quark mass
$m_{ {\rm s}}$, the collective baryon Hamiltonian is no more
SU(3)-symmetric. The octet states are mixed with the higher
representations such as antidecuplet $\overline{{\bf 10}}$ and
eikosiheptaplet ${\bf 27}$~\cite{KimPraGo}.  In the linear order in
$m_{{\rm s}}$ the wave function of a state $B=(Y,I,I_3)$ of spin $S_3$
is given as:
\begin{equation}
\psi _{B,S_3}=(-)^{\frac 12-S_3}\left( \sqrt{8}\,D_{B\,S}^{(8)}+c_B^{(%
\overline{10})}\sqrt{10}\,D_{B\,S}^{(\overline{10})}+c_B^{(27)}\sqrt{27}%
\,D_{B\,S}^{(27)}\right) ,
\end{equation}
where $S=(-1,\frac 12,S_3)$. Mixing parameters $c_B^{({\cal R})}$ can
be found for example in Ref.~\cite{BloPraGo}. They are given as
products of a numerical constant $N_B^{({\cal R})}$ depending on the
quantum numbers of the baryonic state $B$ and dynamical parameter
$c_{{\cal R}}$ depending linearly on $m_{{\rm s}}$ (which we assume to
be 180~MeV) and the model parameter $I_2$, which is responsible for the
splitting between the octet and higher exotic multiplets~\cite
{antidec}.

Analogously to Eq.(\ref{Eq:g1}) one obtains in the $\chi$QSM
diagonal axial-vector coupling constants.
In that case $X$ can take two values: $X=3$ and $X=8$. For $X=0$ (singlet
axial-vector current) we have the following expression \cite{BloPraGo,BPG}:
\begin{equation}
\frac 12\,g_B^{(0)}=\frac 12a_3+\sqrt{3}\,m_{\text{s}}\,(a_5-a_6)\;\langle
B|D_{83}^{(8)}|B\rangle .  \label{Eq:singlet}
\end{equation}

This equation is remarkable, since it provides a link
between an octet and singlet axial-vector current.  It is perhaps the most
important model input in our analysis.  Pure QCD-arguments based the large
$N_c$ expansion~\cite{Man} do not provide such a link.

A remark concerning constants $a_i$ is here in order. Coefficient $a_1$
contains terms which are leading and subleading in the large $N_{{\rm c}}$
expansion. The presence of the subleading terms enhances the numerical value
of $a_1$ calculated in the $\chi$QSM for the self-consistent profile and
makes model predictions {\em e.g.} for $g^{(3)}_{{\rm A}}$ remarkably close
to the experimental data \cite{WaWa,allstars}. This feature, although very
important for the model phenomenology, does not concern us here, since our
procedure is based on fitting all coefficients $a_i$ from the data.
Constants $a_2$ and $a_3$ are both subleading in $1/N_{{\rm c}}$ and come
from the anomalous part of the effective chiral action in Euclidean space.
In the Skyrme model they are related to the Wess-Zumino term. However, in
the simplest version of the Skyrme model (which is based on the
pseudo-scalar mesons only) $a_3 = 0$ identically \cite{BroEllKar}. In the
case of the $\chi$QSM $a_3 \ne 0$ and it provides a link between the SU(3)
octet of axial-vector currents and the singlet current of Eq.(\ref
{Eq:singlet}). It was shown in Ref.\cite{limit} that in the limit of the
artificially large soliton, which corresponds to the ``Skyrme limit'' of the
present model, $a_3/a_1 \rightarrow 0$ in agreement with \cite{BroEllKar}.
On the contrary, for small solitons $g_A^{(0)} \rightarrow 1$ reproducing
the result of the non-relativistic quark model.

So instead of calculating 7 dynamical parameters $a_i (i=1,\cdots,6)$
and $I_2$ (which enters into $c_{\overline{10}}$ and $c_{27}$) within the
$\chi$QSM, we shall fit them from the hyperon semileptonic decays data. It is
convenient to introduce the following set of 7 new parameters:
\[
r=\frac{1}{30} \left( a_1 - \frac{1}{2} a_2 \right),\;\;\; \;\;\;s= \frac{1}{%
60} a_3, \;\;\; x = \frac{1}{540} m_{{\rm s}}\,a_4,\;\;\; y=\frac{1}{90} m_{%
{\rm s}}\,a_5,\;\;\; z=\frac{1}{30}m_{{\rm s}}\, a_6,
\]
\begin{equation}
p= \frac{1}{6}m_{{\rm s}}\, c_{\overline{10}} \left(a_1 + a_2 +\frac{1}{2}
a_3 \right),\;\;\; q=-\frac{1}{90}m_{{\rm s}}\, c_{27} \left(a_1 + 2 a_2 -
\frac{3}{2} a_3 \right) .  \label{Eq:newp}
\end{equation}

Employing this new set of parameters, we can express all possible
semileptonic decays of the octet baryons. Explicit formulae can be found in
the Appendix (see Eq.(\ref{Eq:semilep})). Let us finally note that there is
certain redundancy in Eq.(\ref{Eq:semilep}), namely by redefinition of $q$
and $x$ we can get rid of the variable $p$:
\begin{equation}
x^{\prime}=x- \frac{1}{9} p, \;\;\;\; q^{\prime}=q- \frac{1}{9} p.
\end{equation}
So there are 6 free parameters which have to be fitted from the data.

>From Eq.(\ref{Eq:semilep}), we can easily find that in the chiral limit the
following eight sum rules for $\left({g_1}/{f_1}\right)$ exist:
\begin{eqnarray}
{({\rm n}\rightarrow {\rm p})} = {(\Xi^- \rightarrow \Sigma^0)}, &~~~~& {(%
{\rm n}\rightarrow {\rm p})} = {(\Sigma^{-} \rightarrow {\rm n})} +2 {%
(\Sigma^{+} \rightarrow \Lambda)},  \nonumber \\
{({\rm n}\rightarrow {\rm p})} = \frac{4}{3} {(\Sigma^{+} \rightarrow
\Lambda)} + {(\Xi^{-} \rightarrow \Lambda) }, & & {({\rm n}\rightarrow {\rm p%
})} = {(\Lambda \rightarrow {\rm p})} +\frac{2}{3} {(\Sigma^{+} \rightarrow
\Lambda)},  \nonumber \\
{({\rm n}\rightarrow {\rm p})} = 2 {(\Sigma^{+} \rightarrow \Lambda)} + {%
(\Xi^{-} \rightarrow \Xi^0)}, & & {({\rm n}\rightarrow {\rm p})} = {%
(\Sigma^{-} \rightarrow \Sigma^0)} + {(\Sigma^{+} \rightarrow \Lambda)},
\nonumber \\
{(\Sigma^{+} \rightarrow \Lambda)} = {(\Sigma^{-} \rightarrow \Lambda)}, & &
{(\Xi^0 \rightarrow \Sigma^+)} = {(\Xi^{-} \rightarrow \Sigma^0)}.
\label{srchi}
\end{eqnarray}
Only the first 4 sum rules (\ref{srchi}) contain known decays, and the
accuracy here is not worse than 10 \%.  Apparently the symmetry
breaking of SU(3) has only a small effect on the semileptonic
decays.

With the linear $m_{{\rm s}}$ corrections turned on,
we end up with only four sum rules:
\begin{eqnarray}
{(\Xi^{-} \rightarrow \Sigma^0)} = {(\Xi^0 \rightarrow \Sigma^+)}, &~~~& {%
(\Sigma^{-} \rightarrow \Lambda)} = {(\Sigma^{+} \rightarrow \Lambda)},
\nonumber
\end{eqnarray}
\begin{eqnarray*}
3 {(\Lambda \rightarrow {\rm p})} - 2 {({\rm n}\rightarrow {\rm p})} +2 {%
(\Sigma^{-} \rightarrow {\rm n})} +4 {(\Sigma^{+} \rightarrow \Lambda)} & &
\\
- {(\Xi^{-} \rightarrow \Sigma^0)} +2 {(\Xi^- \rightarrow \Xi^0)} - 2 {%
(\Xi^{-} \rightarrow \Lambda) } & = & 0,
\end{eqnarray*}
\begin{equation}
3 {(\Lambda \rightarrow {\rm p})} - 2 {({\rm n}\rightarrow {\rm p})} - {%
(\Sigma^{-} \rightarrow {\rm n})} +2 {(\Sigma^{+} \rightarrow \Lambda)} -2 {%
(\Xi^{-} \rightarrow \Sigma^0)} +2 {(\Sigma^- \rightarrow \Sigma^0)}= 0.
\label{Eq:sum2}
\end{equation}

However, more experimental data are required to verify Eq.(\ref{Eq:sum2}).

\section{Linking hyperon decays with data on polarized parton distributions}

As we have demonstrated in the preceding Section, the amplitudes of the
hyperon decays are described in the $\chi$QSM by 6 free parameters.  There
are 2 {\em chiral} ones: $r$ and $s$, and
4 proportional to $m_{{\rm s}}$: $x^\prime$, $y$, $z$, and $q^\prime$. Since
there are 6 known hyperon decays, we can
express all model parameters as linear combinations of these decay
constants, and subsequently all quantities of interest can be expressed in
terms of the input amplitudes.  In the following we will use the experimental
values of Refs.~\cite{PDG96,BGHORS}, which
are presented in Table I.

Before doing this, let us, however, observe that there exist two linear
combinations of the decay amplitudes which are free of the $m_{{\rm s}}$
corrections (within the model):
\begin{eqnarray}
A_1 + 2 A_6 & = & -42 r + 6 s \, ,  \nonumber \\
3 A_1 - 8 A_2 - 6 A_3 + 6 A_4 + 6 A_5 & = &90 r + 90 s \, ,  \label{rs0}
\end{eqnarray}
where $A_i$ stand for the decay constants in short-hand notation (see Table
I). Solving Eq.(\ref{rs0}) for $r$ and $s$, we obtain the {\em chiral-limit}
(i.e. with $x^\prime=y=z=q^\prime=0$) expressions for hyperon decays and
integrated
quark densities. The numerical values obtained in this way can be found in
Tables I and II.
Reexpressing $r$ and $s$, $x^\prime$, $y$, $z$ and $q^\prime$
in terms of
the $A_i$'s allows us to write down the integrated quark densities as:
\begin{eqnarray}
\Delta u &=& {\frac{4\,{A_1}}{3}} - {\frac{16\,{A_2}}{9}} - {\frac{4\,{A_3}}{%
3}} + {\frac{4\,{A_4}}{3}} + {\frac{4\,{A_5}}{3}} + {\frac{4\,{A_6}}{3}} \, ,
\nonumber \\
\Delta d &=& {A_1} - {\frac{16\,{A_2}}{9}} - {\frac{4\,{A_3}}{3}} + {\frac{%
4\,{A_4}}{3}} + {\frac{4\,{A_5}}{3}} + {\frac{2\,{A_6}}{3}} \, ,  \nonumber
\\
\Delta s &=& {\frac{2\,{A_1}}{3}} - {\frac{10\,{A_2}}{9}} - {\frac{5\,{A_3}}{%
6}} + {\frac{5\,{A_4}}{6}} + {\frac{5\,{A_5}}{6}} + {\frac{{A_6}}{2}} \, .
\label{uds0}
\end{eqnarray}

The two least known amplitudes $A_5$ and $A_6$ are almost entirely
responsible for the errors quoted in Tables I and II. However, since the
coefficients which enter into Eq.(\ref{uds0}) are not too large, the
absolute errors are relatively small.

In Table II the yet unmeasured hyperon semileptonic decay constants
are listed.  The $\Xi^0\rightarrow \Sigma^+$ channel is particularly
interesting, since its measurement will be soon announced
by the KTeV collaboration~\cite{KTeV}.

Forming linear combinations of the quark densities we obtain the {\em chiral
limit} expressions for $\Gamma_{p,n}$ and $\Delta \Sigma$:
\begin{eqnarray}
\Gamma_p &=& 7\,{A_1} - 10\,{A_2} - {\frac{15\,{A_3}}{2}} + {\frac{15\,{A_4}%
}{2}} + {\frac{15\,{A_5}}{2}} + {\frac{13\,{A_6}}{2}} \, ,  \nonumber \\
\Gamma_n &=& 6\,{A_1} - 10\,{A_2} - {\frac{15\,{A_3}}{2}} + {\frac{15\,{A_4}%
}{2}} + {\frac{15\,{A_5}}{2}} + {\frac{9\,{A_6}}{2}} \, ,  \nonumber \\
\Delta \Sigma &=& 3\,{A_1} - {\frac{14\,{A_2}}{3}} - {\frac{7\,{A_3}}{2}} + {%
\frac{7\,{A_4}}{2}} + {\frac{7\,{A_5}}{2}} + {\frac{5\,{A_6}}{2}} \, .
\end{eqnarray}
The numerical values together with the error bars are listed in Table II.

The full expressions are obtained by solving the
remaining 4 equations for $m_{{\rm s}}$ dependent parameters
$x^\prime$, $y$, $z$ and $q^\prime$.  Also in this case
we are able to link integrated quark densities $\Delta q$ to the hyperon
decays:
\begin{eqnarray}
\Delta u &=& {\frac{8\,{A_2}}{9}} + {\frac{5\,{A_3}}{3}} + {\frac{7\,{A_4}}{3%
}} + {\frac{{A_5}}{3}} - {\frac{{A_6}}{3}}\, ,  \nonumber \\
\Delta d &=& -{A_1} + {\frac{8\,{A_2}}{9}} + {\frac{5\,{A_3}}{3}} + {\frac{%
7\,{A_4}}{3}} + {\frac{{A_5}}{3}} - {\frac{{A_6}}{3}} \, ,  \nonumber \\
\Delta s &=& {\frac{15\,{A_1}}{4}} - {\frac{101\,{A_2}}{18}} - {\frac{289\,{%
A_3}}{48}} + {\frac{13\,{A_4}}{48}} + {\frac{43\,{A_5}}{48}} + {\frac{149\,{%
A_6}}{48}} \, .
\end{eqnarray}
It is interesting to observe that the amplitudes $A_5$ and in particular $%
A_6 $ come with relatively large weight in the expression for $\Delta s$,
whereas $\Delta u$ and $\Delta d$ are much less affected by the
relatively large experimental error of these two decays. This is
explicitly seen in Fig.1,
where we plot the central values and error bars of $\Delta q$'s. In the same
figure we draw central values and errors of $\Delta q$'s in the {\em chiral
limit} as given by Eq.(\ref{uds0}). To guide the eye we have restored the
linear dependence on the symmetry breaking $m_{{\rm s}}$ corrections
assuming $m_{\rm s}=180$~MeV, as done in Ref.\cite{KimPraGo}.

We can first see that our results in the chiral limit correspond to typical
SU(3)-symmetric values: $F\approx 0.50$ and $D\approx 0.77$. However, the
result for individual integrated quark densities,
where the model prediction for the
singlet current $g_{{\rm A}}^{(0)}$ plays a role, are beyond the typical
SU(3) symmetry values. Only when the chiral symmetry breaking is taken into
account the central values for $\Delta q$'s are shifted towards the
"standard" values. Unfortunately the error of $\Delta s$ becomes 7 times
larger than the one of $\Delta u$ or $\Delta d$, so that at this stage we
are not able to make any firm conclusion concerning the value of $\Delta s$.

It is perhaps more interesting to look directly at the combinations relevant
for the polarized scattering experiments, which take the following form:
\begin{eqnarray}
\Gamma_p &=& {\frac{11\,{A_1}}{4}} - {\frac{7\,{A_2}}{6}} + {\frac{37\,{A_3}%
}{16}} + {\frac{191\,{A_4}}{16}} + {\frac{41\,{A_5}}{16}} + {\frac{23\,{A_6}%
}{16}} \, ,  \nonumber \\
\Gamma_n &=& {\frac{-{A_1}}{4}} - {\frac{7\,{A_2}}{6}} + {\frac{37\,{A_3}}{16%
}} + {\frac{191\,{A_4}}{16}} + {\frac{41\,{A_5}}{16}} + {\frac{23\,{A_6}}{16}%
}\, ,  \nonumber \\
\Delta \Sigma &=& {\frac{11\,{A_1}}{4}} - {\frac{23\,{A_2}}{6}} - {\frac{43\,%
{A_3}}{16}} + {\frac{79\,{A_4}}{16}} + {\frac{25\,{A_5}}{16}} + {\frac{39\,{%
A_6}}{16}}\, .  \label{GDSfull}
\end{eqnarray}
In Fig.2 we plot $\Gamma_{p,n}$ and $\Delta \Sigma$ both for the chiral
symmetry fit and for the full fit of Eq.(\ref{GDSfull}), together with
experimental data for the proton and neutron. Again,
to guide the eye we have restored the
linear dependence of the symmetry breaking $m_{{\rm s}}$ corrections.
We see that despite the large
uncertainty of $\Delta s$, we get reasonable values for $\Gamma_p$ and $%
\Gamma_n$. Somewhat unexpectedly we see, that $\Delta \Sigma$ is almost
independent of the chiral symmetry breaking\footnote{Similar behavior
has been observed in Ref.\cite{LichtLip}.} and stays within the range
$0.1 \rightarrow 1.1$, if the errors of the hyperon decays are taken
into account. $75 \% $
of the experimental error of $\Delta \Sigma$ comes from the two least known
hyperon decays $\Xi^- \rightarrow \Lambda,\, \Sigma^0 $ (corresponding to $%
A_5$ and $A_6$).

It is interesting to see how $\Delta\Sigma$ and $\Delta s$ are correlated.
To this end, instead of using two last hyperon decays $A_5$ and $A_6$ as
input, we use the experimental value for $\Gamma_p$ as given by Eq.(\ref
{Gampval}) and $\Delta\Sigma$, which we vary in the range from 0 to 1. In
Fig.3 we plot our prediction for the two amplitudes $A_5$ and $A_6$ (solid
lines), together with the experimental error bands for these two decays. It
is clearly seen from Fig.3 that the allowed region for $\Delta\Sigma$, in
which the theoretical prediction falls within the experimental error bars
amounts to $\Delta\Sigma=0.20\rightarrow 0.45$.

In Fig.4 we plot the variation of $\Delta q$'s with respect to $\Delta
\Sigma $ (with $\Gamma_p$ fixed by Eq.(\ref{Gampval})).
We see that $\Delta u$ and $\Delta d$ are relatively stable,
whereas $\Delta s$ exhibits rather strong dependence on $\Delta \Sigma$.
Within the allowed region $0.20 < \Delta \Sigma <0.45 $ strange quark
density $\Delta s$ varies between $-0.12$ and $0.30$. Interestingly, in the
central region around $\Delta\Sigma\approx 0.30$ strange quark density
vanishes in accordance with an intuitive assumption of Ellis and Jaffe~\cite
{EllisJaffe}.

Identical behavior\footnote{Note that authors of Ref.\cite{LichtLip}
use a slightly different value for $I_p$ and include higher order QCD
corrections.} (shown in Fig.4 by a dash-dotted line) was obtained by
Lichtenstadt and Lipkin in an analysis of
the hyperon decays in which no model for $\Delta \Sigma$ has been
used \cite{LichtLip}. Indeed (assuming only the first order QCD corrections),
the identity:
\begin{equation}
\Delta \Sigma = \frac{1}{2} \Gamma_p
- \frac{1}{4} \left(3 g_{\rm A}^{(3)}+\sqrt{3} g_{\rm A}^{(8)} \right)\ .
\label{nomodel}
\end{equation}
allows one to calculate $\Delta \Sigma$ in terms of $ g_{\rm A}^{(8)} $
(or equivalently $\Delta s$) by using $ g_{\rm A}^{(3)}=1.257$ and
$\Gamma_p$ as an additional input. In  the $\chi$QSM and also in large $N_c$
QCD one can express $ g_{\rm A}^{(8)}$ in terms of the known
hyperon semileptonic decays:
\begin{equation}
\left(3 g_{\rm A}^{(3)}+\sqrt{3} g_{\rm A}^{(8)} \right)=
\frac{1}{8}
\left(-44\,{  A_1} + 104\,{  A_2} + 123\,{  A_3} + 33\,{  A_4} -
     9\,{  A_5} - 55\,{  A_6} \right) \ .
\label{nomodel1}
\end{equation}
Equation (\ref{nomodel1}) gives $\Delta \Sigma= 0.46\pm 0.31$,
remarkably close to the $\chi$QSM prediction in which model formula for
$\Delta \Sigma$ is used. This is, in our opinion, another strong
argument in support for the model formula for $  g_{\rm A}^{(8)} $.

\section{Summary}

In this paper we studied the influence of the SU(3) symmetry breaking
in semileptonic hyperon decays on the determination of the integrated
polarized quark densities $\Delta q$. Using the Chiral Quark Soliton
Model we have obtained a satisfactory parametrization of all available
experimental data on semileptonic decays. In this respect our analysis
is identical to the large $N_{\rm c}$ QCD analysis of Ref.\cite{Man}.
Using 6 known hyperon decays we have predicted $g_1/f_1$ for the decays
not yet measured.

The new ingredient of our analysis consists in using the model formula
for the singlet axial-vector current in order to make contact with the
high energy polarization experiments. We have argued that our model
interpolates between the quark model (the small soliton limit) and the
Skyrme model (large soliton limit) \cite{DiaMog} reproducing the value
of $\Delta \Sigma$ in these two limiting cases \cite{limit,charge}.
This unique feature, and also the numerical agreement with the analysis
of Ref.\cite{LichtLip} as discussed at the end of the last Section,
make us believe that our approach contains all necessary physics needed
to analyze the symmetry breaking not only for the octet axial-vector
currents, but also in the case of the singlet one.

The model contains 6 free parameters which can be fixed by 6 known
hyperon decays.  Unfortunately $g_1/f_1$ for the two known decays of
$\Xi^-$ have large experimental errors, which influence our predictions
for $\Delta q$.  Our strategy was very simple: using model
parametrization we expressed $\Delta q$'s, $\Gamma_{p,n}$ and $\Delta
\Sigma$ in terms of the six known hyperon decays. Errors were added in
quadrature.

First observation which should be made is that we reproduce
$\Gamma_{p,n}$ as measured in deep inelastic scattering.  We obtain
$\Delta u=0.72\pm0.07$ and $\Delta d=-0.54\pm0.07$, however, $\Delta s$
is practically undetermined being equal to $0.33\pm0.51$. This large
error is entirely due to the experimental errors of the $\Xi^-$ decays,
which also make $\Delta
\Sigma$ to lie between 0.1 and 0.9.

There are two points which have to be stressed here.  Our fit respects
chiral symmetry in a sense that the leading order parameters $r$ and
$s$ (or equivalently $F$ and $D$) are fitted to the linear combinations
of the hyperon decays which are free from $m_{\rm s}$ corrections. Had
we used this SU(3) symmetric parametrization as given by Eq.(\ref{rs0})
we would not be able to reproduce (as far as the central values are
concerned) $\Gamma_{p,n}$. With $m_{\rm s}$ corrections turned on we
hit experimental values for $\Gamma_{p,n}$, however, as stated above,
the value of $\Delta \Sigma$ is practically undetermined, due the the
experimental error of $\Xi^-$ decays. Therefore to confirm or
invalidate our analysis  it is of utmost importance to have better data
for these decays. Since we predict that $(\Xi^- \rightarrow \Sigma^0)=
(\Xi^0 \rightarrow \Sigma^+)$ the forthcoming experimental result for
the latter decay \cite{KTeV} will provide a test of our approach.
If the future data on this and on other decays will disagree with the
predictions of our analysis (in which dynamical quantities are {\em
fitted} to the existing data rather than {\em calculated} in the
model), that would also mean that the model (with dynamical quantities
{\em calculated}) fails for these particular observables. It would be
then the signal for the model builders that presumably there were some
physical effects  which had been not  included in the present version
of the model.

Interestingly, if we use $\Gamma_p$ and $\Delta \Sigma$ as an input
instead of the $\Xi^-$ decays, we see very strong correlation between
$\Delta \Sigma$ and $\Delta s$, whereas $\Delta u$ and $\Delta d$ are
basically $\Delta \Sigma$ independent. This behavior has been also
observed in Ref.\cite{LichtLip}.

Our analysis shows clearly that if one wants to link the low-energy
hyperon semileptonic decays with high-energy polarized experiments, one
cannot neglect SU(3) symmetry breaking for the former. In this respect
our conclusions agree with Refs.\cite{EhrSch,SavWal}. Similarly to
Ref.\cite{EhrSch} we see that $\Delta s=0$ is not ruled out by present
experiments.  Therefore the results for $\Delta s$ and $\Delta \Sigma$
which are based on the exact SU(3) symmetry are in our opinion
premature.  The  meaningful results for these  2 quantities can be
obtained only if the experimental errors for the $\Xi^-$ decays are
reduced.

\section*{Acknowledgments}

This work has partly been supported by the BMBF, the DFG and the
COSY--Project (J\" ulich).  We are grateful to M.V. Polyakov for
fruitful discussions.
H.-Ch.K. and M.P. thank P.V.  Pobylitsa for critical comments.  M.P.
thanks M. Karliner for his comment
concerning Ref.\cite{LichtLip}.  H.-Ch.K. has been
supported by Pusan National University Research Grant. M.P. has been
supported by Polish grant {PB~2~P03B~019~17}.

\section*{appendix}

In this Appendix we quote the formulae used in the fits.
Semileptonic decay constants are parametrized as follows:
\begin{eqnarray}
A_1\;=\;
\left({g_1}/{f_1}\right)^{({\rm n}\rightarrow {\rm p})} & = &
-14 r + 2 s - 44 x - 20 y - 4 z-4 p + 8 q,  \nonumber \\
A_2\;=\;
\left({g_1}/{f_1}\right)^{(\Sigma^{+}\rightarrow \Lambda)} & = &
- 9r - 3s - 42x - 6y -3p + 15q,  \nonumber \\
A_3\;=\;
\left({g_1}/{f_1}\right)^{(\Lambda \rightarrow {\rm p})} & = &
- 8r + 4s + 24x - 2z+ 2p - 6q,  \nonumber \\
A_4\;=\;
\left({g_1}/{f_1}\right)^{(\Sigma^{-} \rightarrow {\rm n})} & = &
4r + 8s -4x - 4y + 2z+4q,  \nonumber \\
A_5\;=\;
\left({g_1}/{f_1}\right)^{(\Xi^{-} \rightarrow \Lambda) } & = &
- 2r + 6s - 6x + 6y - 2 z + 6q,  \nonumber \\
A_6\;=\;
\left({g_1}/{f_1}\right)^{(\Xi^{-} \rightarrow \Sigma^0)} & = &
- 14r + 2s + 22x + 10 y + 2 z +2p - 4q,  \nonumber \\
\left({g_1}/{f_1}\right)^{(\Sigma^{-} \rightarrow \Lambda)} & = &
- 9 r - 3 s - 42x - 6y- 3p + 15q,  \nonumber \\
\left({g_1}/{f_1}\right)^{(\Sigma^- \rightarrow \Sigma^0)} & = &
- 5r + 5s - 18x - 6y + 2 z- 2p,  \nonumber \\
\left({g_1}/{f_1}\right)^{(\Xi^- \rightarrow \Xi^0)} & = &
4r + 8s + 8x + 8y - 4 z - 8q,  \nonumber \\
\left({g_1}/{f_1}\right)^{(\Xi^0 \rightarrow \Sigma^+)} & = &
- 14r + 2 s + 22 x + 10 y + 2 z+ 2p - 4 q .  \label{Eq:semilep}
\end{eqnarray}

The U(1) and SU(3) axial-vector constants $g_A^{(0,3,8)}$
can be also expressed in
terms of the new set of parameters (\ref{Eq:newp}). For the singlet
axial-vector constant, we have
\begin{equation}
g_A^{(0)}=60s-18y+6z,\label{Eq:singlet1}
\end{equation}
for the triplet one\footnote{%
Triplet $g_{A}^{(3)}$'s are proportional to $I_3$, formulae in Eq.(\ref
{Eq:triplet}) correspond to the highest isospin state.}:
\begin{equation}
g_A^{(3)}=-14r+2s-44x-20y-4z-4p+8q, \label{Eq:triplet}
\end{equation}
and for the octet one, we get:
\begin{equation}
g_A^{(8)}=\sqrt{3}(-2r+6s+12x+4p+24q).  \label{Eq:octet}
\end{equation}

\newpage


\vfill

\newpage

\begin{center}
{\Large {\bf Figure Captions}}
\end{center}

\noindent
{\bf Fig. 1}: $\Delta q$ as a function of the strange quark mass $m_s$.
While the $\Delta u$ and $\Delta d$ have less uncertainties as the $m_s$
increases, the uncertainty of $\Delta s$ becomes larger, as the $m_s$
increases. \vspace{0.8cm}

\noindent
{\bf Fig. 2}: $\Gamma_ {p,n}$ and $\Delta \Sigma$ as functions of $m_s$.
While the uncertainty of $\Gamma_ {p,n}$ decreases, as the $m_s$ increases,
the error of the $\Delta \Sigma$ remains constant. The error bars denote the
experimental data for the $\Gamma_ {p,n}$. \vspace{0.8cm}

\noindent
{\bf Fig. 3}: $A_5$ (lower line) and $A_6$ (upper line) as functions of $%
\Delta \Sigma$. \vspace{0.8cm}

\noindent
{\bf Fig. 4}: $\Delta q$'s as functions of $\Delta \Sigma$.  Dash-dotted
line below $\Delta s$ corresponds to the result of Ref.~\cite{LichtLip}.

\vfill

\newpage

\begin{center}
{\Large {\bf Figures}}
\end{center}

\vspace{1.6cm} \centerline{\epsfysize=2.7in\epsffile{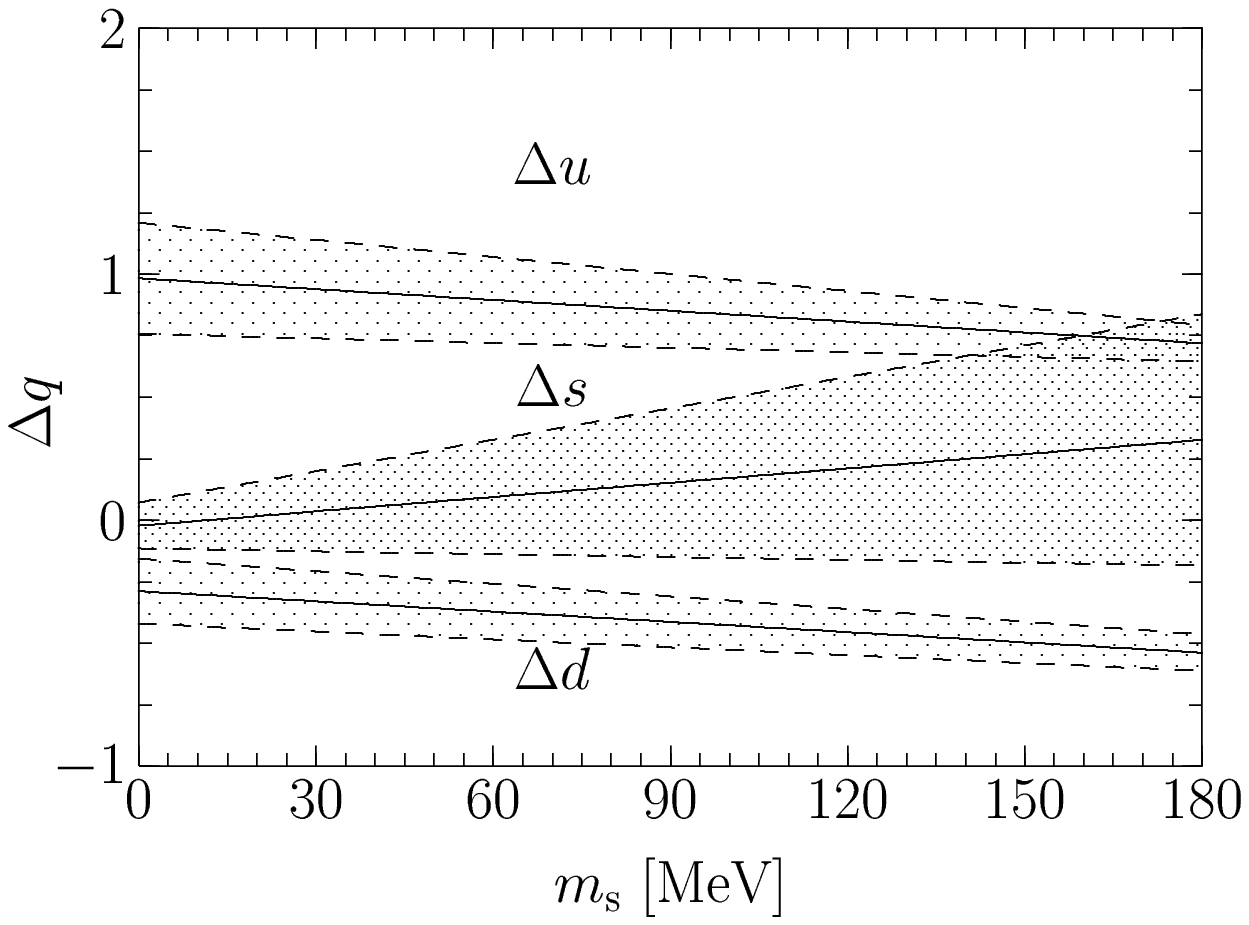}}\vskip4pt
\noindent

\begin{center}
{\bf Figure 1}
\end{center}

\vspace{1.6cm} \centerline{\epsfysize=2.7in\epsffile{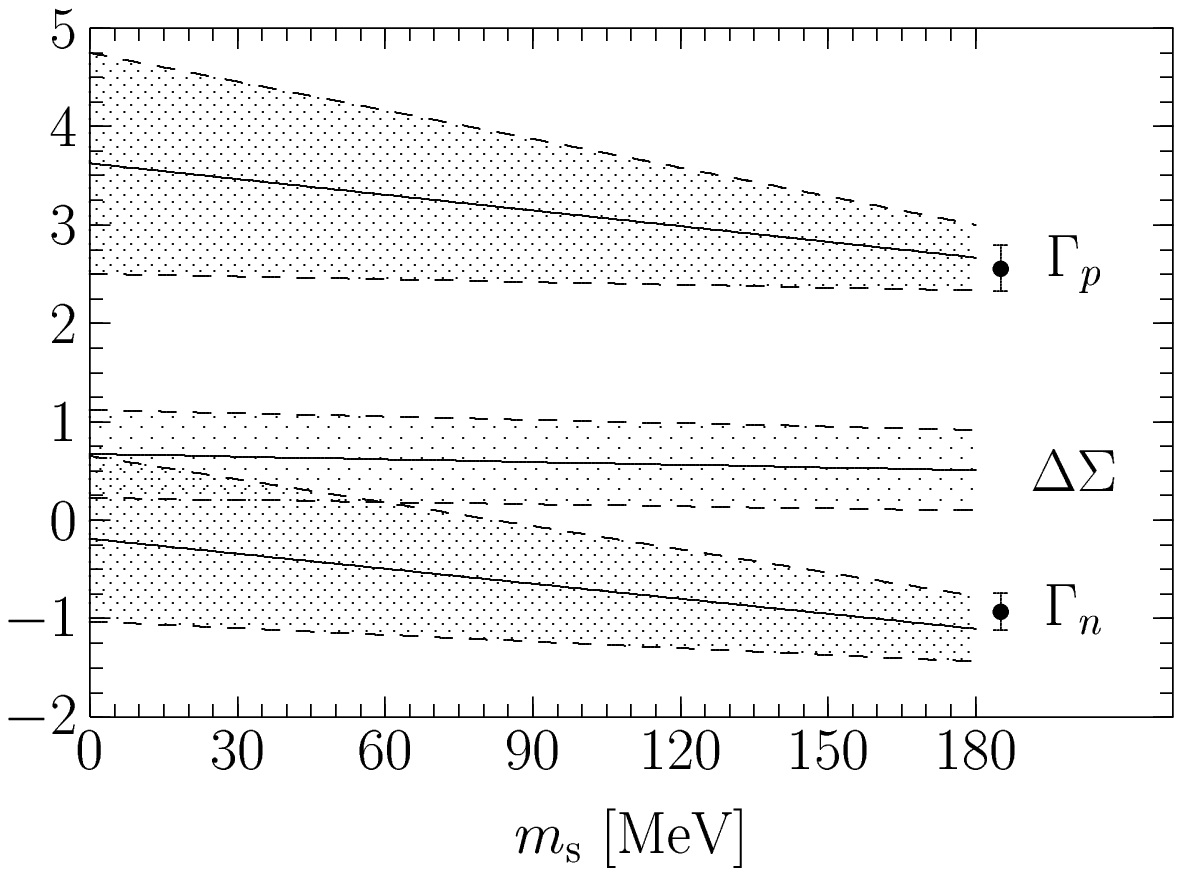}}\vskip4pt
\noindent

\begin{center}
{\bf Figure 2}
\end{center}

\vspace{1.6cm} \centerline{\epsfysize=2.7in\epsffile{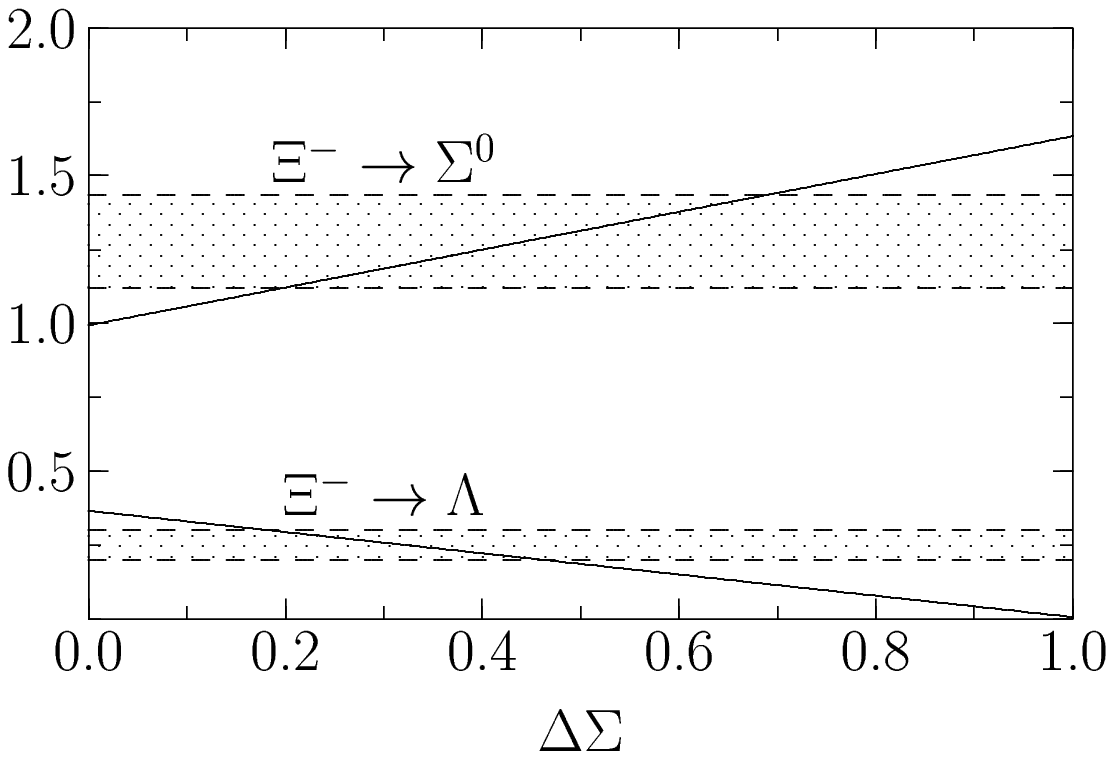}}\vskip4pt
\noindent

\begin{center}
{\bf Figure 3}
\end{center}

\vspace{1.6cm} \centerline{\epsfysize=2.7in\epsffile{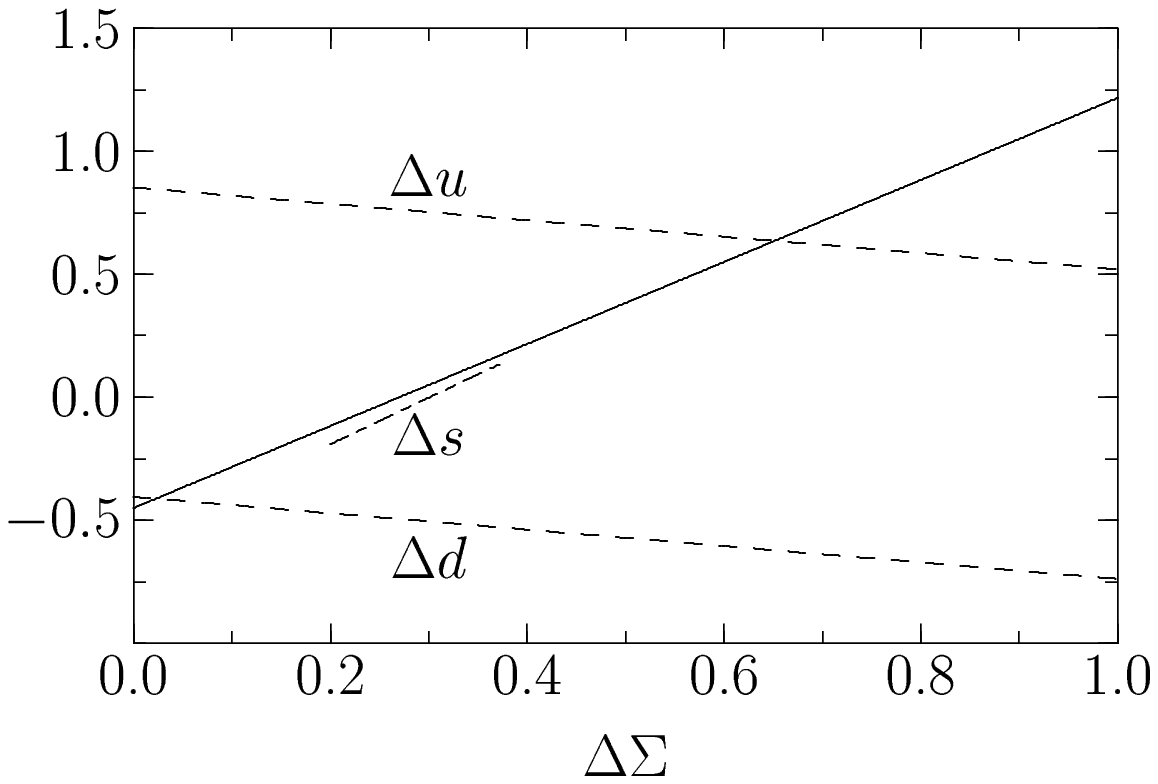}}\vskip4pt
\noindent

\begin{center}
{\bf Figure 4}
\end{center}

\newpage

\begin{table}
\caption{The parameters $r, \ldots, q^\prime$  fixed to the 
experimental data of the semileptonic decays
\protect\cite{PDG96,BGHORS} $A_1$ -- $A_6$. 
The entries for $A_1$ -- $A_6$ for the full fit (last column)
correspond to the experimental data.}
\begin{tabular}{cccc}
& & chiral limit  & with $m_{\rm s}$            \\ \hline 
& $r$ & $-0.0892 $  & $-0.0892             $      \\    
& $s$ & $ 0.0113 $  & $ 0.0113              $	   \\ 
& $x^{\prime}$ & 
      $ 0~~~~  $  & $-0.0055              $	   \\
& $y$ & $ 0~~~~  $  & $ 0.0080              $	   \\
& $z$ & $ 0~~~~  $  & $-0.0038              $	   \\
& $q^{\prime}$ & 
      $ 0~~~~  $  & $ -0.0140              $	   \\
\hline
$A_1$ & $\left({g_1}/{f_1}\right)^{n\rightarrow p}$        
      &$  1.271\pm 0.11$ & $  1.2573\pm 0.0028$\\
$A_2$ & $\left({g_1}/{f_1}\right)^{\Sigma^+\rightarrow \Lambda}$ 
      &$  0.769\pm 0.04$ & $0.742 \pm 0.018~~ $\\
$A_3$ & $\left({g_1}/{f_1}\right)^{\Lambda\rightarrow p}$        
      &$  0.758\pm 0.08$ & $  0.718 \pm 0.015~~ $\\
$A_4$ & $\left({g_1}/{f_1}\right)^{\Sigma^-\rightarrow n}$       
      &$ -0.267\pm0.04$ & $ -0.340 \pm 0.017~~ $\\
$A_5$ & $\left({g_1}/{f_1}\right)^{\Xi^-\rightarrow \Lambda}$    
      &$  0.246\pm 0.07$ & $  0.25  \pm 0.05~~~  $\\
$A_6$ & $\left({g_1}/{f_1}\right)^{\Xi^-\rightarrow \Sigma^0}$   
      &$  1.271\pm 0.11$ & $  1.278 \pm 0.158~~ $\\ 
\end{tabular}
\vspace{1cm}
\caption{The predictions for yet unmeasured decays, 
integrated quark densities $\Delta q$ and $\Gamma_{p,n}$ and 
$\Delta\Sigma$.}
\begin{tabular}{cccc}
& & chiral limit  & with $m_{\rm s}$            \\ \hline 
      & $\left({g_1}/{f_1}\right)^{\Sigma^-\rightarrow \Lambda}$  
      &$  0.769\pm0.04$ & $  0.742\pm 0.02$ \\ 
      & $\left({g_1}/{f_1}\right)^{\Sigma^-\rightarrow \Sigma^0}$ 
      &$  0.502\pm 0.07$  & $  0.546\pm 0.16$ \\
      & $\left({g_1}/{f_1}\right)^{\Xi^-\rightarrow \Xi^0}$       
      & $-0.267\pm 0.04$  & $ -0.12\pm 0.12$ \\
      & $\left({g_1}/{f_1}\right)^{\Xi^0\rightarrow \Sigma^+}$    
      &$  1.271 \pm 0.11$ & $ 1.278\pm 0.16$ \\ \hline
     & $\Delta u$   &$ 0.98\pm 0.23$ &$ 0.72 \pm 0.07$   \\
     & $\Delta d$  &$-0.29\pm 0.13$ &$-0.54\pm 0.07$      \\
     & $\Delta s$  &$-0.02\pm 0.09$ &$0.33\pm 0.51$       \\ \hline
     & $\Gamma_p$  &$3.63 \pm 1.12$  & $ 2.67 \pm 0.33 $   \\
     & $\Gamma_n$  &$-0.19 \pm 0.84$ & $-1.10 \pm 0.33 $  \\
     & $\Delta\Sigma$ 
                   &$ 0.68 \pm 0.44$ & $ 0.51 \pm 0.41 $
\end{tabular}

\end{table}

\end{document}